\documentstyle[preprint,eqsecnum,aps]{revtex}

\begin{document}

\draft

\preprint{LA-UR-95-1648}
%\preprint{TPR-95-10}

\title{
  Extracting source parameters from gaussian fits \\
  to two-particle correlations
}

\author
  {Scott Chapman and J. Rayford Nix}

\address{
  Theoretical Division, Los Alamos National Laboratory,
  Los Alamos, NM  87545, USA
}

\author
{Ulrich Heinz}

\address
 {Institut f\"ur Theoretische Physik, Universit\"at Regensburg, \\
  D-93040 Regensburg, Germany}

\date{August 9, 1995}

\maketitle

\begin{abstract}
Using a quadratic saddle-point approximation, we show how information
about a particle-emitting source can be extracted from gaussian fits
to two-particle correlation data.  Although the formalism is
completely general, extraction of the relevant parameters is much
simpler for sources within an interesting class of azimuthally
symmetric models.  After discussing the standard fitting procedure, we
introduce a new gaussian fitting procedure which is an azimuthally
symmetric generalization of the Yano-Koonin formalism for spherically
symmetric sources. This new fitting procedure has the advantage that
in addition to being able to measure source parameters in a fixed
frame or the longitudinally co-moving system, it can also measure
these parameters in the local rest frame of the source.
\end{abstract}
\pacs{
PACS:
25.70.Pq}

\narrowtext

%%%%%%%%%%%%%%%%%%%%%%%%%%%%%%%%%%%%%%%%%%%%%%%%%%%%%%%%%%%%%%%%%
\section{Introduction}\label{s1}
%%%%%%%%%%%%%%%%%%%%%%%%%%%%%%%%%%%%%%%%%%%%%%%%%%%%%%%%%%%%%%%%%

Recently quite a bit of work has been done in trying to determine
which attributes of the hadronic source formed in high-energy particle
or heavy-ion collisions can be determined by measuring the
Hanbury-Brown--Twiss (HBT) correlations of identical emitted
particles.  Usually the experimental correlation function is fit with
a gaussian in some components of the four-momentum difference
$q=p_1-p_2$ \cite{hbt,na35alb,fixed,both,lcms}. The parameters of such
a fit (called correlation radii) are then often compared to some
simple analytic model in order to get an idea of what is being
measured. For example, for a static gaussian source, the duration of
emission time is directly proportional to the difference of the
squares of the correlation radii which are parallel to (``out'') and
perpendicular to (``side'') the transverse component of the total pair
momentum.  However, this result is not true if for example the actual
source contains any $z$-$t$, $z$-$x$, or $x$-$t$ correlations, such as
would be caused for quickly expanding sources.

The purpose of this paper is to determine exactly what features of the
source are actually being measured by experimental correlation data.
To do this, we use a quadratic saddle-point approximation \cite
{sinyu1,akke,chap2} to a general source function in order to derive a
completely Lorentz covariant expression for the two-particle
correlation function which can be applied to a wide range of analytic
models. In this approximation, 10 ${\bf K}$-dependent parameters are
needed to describe a general source, where ${\bf
K}={\textstyle\frac{1}{2}}({\bf p}_1+{\bf p}_2)$ is the average
momentum of the two particles. The origin of these 10 parameters can
be understood by noting that for each value of ${\bf K}$, our
approximation is mathematically equivalent to a gaussian ellipsoid
described by three Euler angles of orientation, three components of
the velocity of the local rest frame, three spatial extensions, and
one temporal extension. As we will show, however, only six ${\bf
K}$-dependent parameters (not including the chaoticity parameter
$\lambda$) can be measured by making a gaussian fit in $q$ to the
correlation function.

Furthermore, a source function which is azimuthally symmetric in
coordinate space will not in general be azimuthally symmetric in
momentum space, since the direction defined by $K_\perp$ breaks this
symmetry. As a result, the ${\bf K}$-dependent ellipsoid equivalent to
such a source still requires one Euler angle, two velocity components,
three spatial extensions, and one temporal extension in order to
describe it. In this case, only four ${\bf K}$-dependent parameters
can be determined by making a gaussian fit in $q$. Since for arbitrary
$K_\perp$ the number of source parameters exceeds the number of
gaussian-fit parameters, some definite model must be used in order to
interpret how the latter depend on the former.

For pairs with $K_\perp=0$, however, only four ${\bf K}$-dependent
parameters are needed to describe the source (one velocity component,
two spatial extensions and one temporal extension), so the four
parameters measured in a gaussian fit provide enough information to
unambiguously determine all of these source parameters. Although in
practice it is very difficult to measure a pair whose total transverse
momentum vanishes, there is an interesting class of models in which
the $K_\perp =0$ simplifications persist for nonzero values of
$K_\perp$. In Sec.~5 we study a hydrodynamical model similar to ones
which have been used to fit one-particle distributions from heavy-ion
collisions at the Brookhaven Alternating Gradient Synchrotron (AGS)
and the CERN Super Proton Synchrotron (SPS).  Within the context of
this model, one would expect the simplifications in the extraction of
source parameters to occur for pairs whose average energy in the
measurement frame is less than the freezeout temperature divided by
the square of the transverse flow velocity (about 560 MeV for Si+Au at
the AGS or S+Pb at the SPS).

If the correlation function for a model in the afore-mentioned class
is fit to a gaussian in the spatial components of $q$, then in order
to extract the relevant source parameters it is crucial to pick
beforehand the correct longitudinal reference frame for the
measurement.  If, on the other hand, a fit is made to a
generalization of the Yano-Koonin formalism \cite{yano,zajc}, the
correlation radii will automatically measure the relevant source
parameters even if the wrong frame is chosen for measurement.  For
example, for a finite longitudinally expanding source, the new
formalism allows measurement of the source parameters in the local
rest frame of the fluid, whereas fixed and longitudinally co-moving
system (LCMS) radii measure the source in different frames.

%%%%%%%%%%%%%%%%%%%%%%%%%%%%%%%%%%%%%%%%%%%%%%%%%%%%%%%%%%%%%%%%%
\section{Lorentz Covariant Correlation Function}\label{s2}
%%%%%%%%%%%%%%%%%%%%%%%%%%%%%%%%%%%%%%%%%%%%%%%%%%%%%%%%%%%%%%%%%

For the two-particle correlation function, we use the well-established
theoretical approximation
\cite{chap2,pratt,chap1}
 \begin{equation}
   C({\bf q}, {\bf K}) \simeq 1 \pm \frac{|\int d^4x\,
   S(x,K)\,e^{iq{\cdot}x}|^2} {|\int d^4x\, S(x,K)|^2}\;,
 \label{1.0}
 \end{equation}
where ${\bf q}={\bf p}_1-{\bf p_2}$, $q_0 = E_1-E_2$, ${\bf K} =
{\textstyle\frac{1}{2}}({\bf p}_1+{\bf p}_2)$, and $K_0 = E_K =
\sqrt{m^2+|{\bf K}|^2}$.  The plus sign is to be used for boson pairs
and the minus sign for fermion pairs.  The labeling of particles $1$
and $2$ is defined such that $q_1=q_x$ is always positive. In this
way, pairs with positive $q_2=q_y$ and/or $q_3=q_z$ are physically
distinct from those with negative $q_2$ and/or $q_3$. The $S(x,K)$ in
Eq.~(\ref{1.0}) is a function which describes the phase-space density
of the emitting source.

The spacetime saddle point $\bar{x}({\bf K})$ of the emission
function $S(x,K)$ is defined via the four equations
\cite{sinyu1,akke,chap2}
 \begin{equation}
   \frac{\partial}{\partial x_\mu}{\rm ln} S(x,K) \Bigr|_{\bar{x}} = 0\;,
 \label{1.1}
\end{equation}
where $\mu = \{0,1,2,3\}$.  The saddle point is that point in
spacetime which has the maximum probability of emitting a particle
with momentum ${\bf K}$. A quadratic saddle-point approximation for
$S(x,K)$ then yields
\begin{equation}
   S(x,K) \simeq S(\bar{x},K)\exp\left[-{\textstyle\frac{1}{2}}
   (x-\bar{x})^\mu(x-\bar{x})^\nu B_{\mu\nu}({\bf K})\right]\;,
 \label{1.2}
 \end{equation}
where the symmetric curvature tensor $B_{\mu\nu}$ is given by
 \begin{equation}
   B_{\mu\nu}({\bf K})
  =-\partial_\mu\partial_\nu {\rm ln} S(x,K)\Bigr|_{\bar{x}}\;.
 \label{1.3}
 \end{equation}
We define the curvature radius in the $\mu$th direction by
\cite{sinyu1,akke}
 \begin{equation}
   \lambda_\mu({\bf K}) = \left[B_{\mu\mu}({\bf K})\right]^{-1/2}\;.
 \label{2.3}
 \end{equation}
Note that since $B_{\mu\nu}({\bf K})$ is symmetric, it will in general
have 10 independent components. From the form of Eq.~(\ref{1.2}) it is
seen that the saddle-point approximation is mathematically equivalent
to an ellipsoid described by the 10 ${\bf K}$-dependent parameters
mentioned in the Introduction. As long as the saddle point
$\bar{x}({\bf K})$ is unique, knowledge of the 10 functions
$B_{\mu\nu}({\bf K})$ is in most practical situations sufficient for a
complete characterization of the source.

It is convenient to define \cite{chap2,bert1,chap3} the following
${\bf K}$-dependent average of an arbitrary spacetime function
$\xi(x)$ with the source density $S(x,K)$:
\begin{equation}
   \langle \xi \rangle \equiv \langle \xi(x) \rangle ({\bf K})
   = {\int d^4x \, \xi(x)\, S(x,K) \over \int d^4x \, S(x,K) }\; .
 \label{1.6a}
\end{equation}
Using this notation, the correlation function (\ref{1.0}) can be
compactly written as
\begin{equation}
   C({\bf q}, {\bf K}) \simeq 1 \pm \left\vert \langle e^{iq{\cdot}x}
   \rangle ({\bf K}) \right\vert^2\; .
 \label{1.6b}
 \end{equation}
Furthermore, within the saddle-point approximation (\ref{1.2}), the
following relations hold:
 \begin{equation}
    \langle x_\mu \rangle = \bar x_\mu({\bf K})\; ,
 \label{rem1}
 \end{equation}
 \begin{equation}
    \langle x_\mu x_\nu \rangle  - \langle x_\mu \rangle \,
    \langle x_\nu \rangle = (B^{-1})_{\mu\nu}({\bf K})\; .
 \label{rem2}
\end{equation}
The saddle point is thus the average spacetime point from which
particle pairs with momentum ${\bf K}$ are emitted, and the components
of the inverse of the curvature tensor $(B^{-1})_{\mu\nu}({\bf K})$
give the spacetime correlations of the source.  The four diagonal
elements
\begin{equation}
    (B^{-1})_{\mu\mu}({\bf K})  =
        \langle x_\mu^2 \rangle  - \langle x_\mu \rangle^2
 \label{rem3}
\end{equation}
can be understood as the squares of the {\em lengths of homogeneity}
of the source as seen by pairs with momentum ${\bf K}$. It should be
noted that the homogeneity lengths agree with the curvature radii
(\ref{2.3}) only if the curvature tensor $B_{\mu\nu}$ is diagonal.
This was implicitly assumed by the authors of \cite{sinyu1,akke}, who
first introduced the name ``homogeneity length" but used it for the
curvature radii (\ref{2.3}) of the source near the saddle point.

Within the approximations of Eqs.~(\ref{1.0}) and (\ref{1.2}),
calculation of a general correlation function is straightforward,
yielding
\begin{equation}
   C({\bf q},{\bf K})
  = 1 \pm \exp\left[-q^\mu q^\nu (B^{-1})_{\mu\nu}\right]\, .
 \label{1.4}
\end{equation}
{}From Eq.~(\ref{rem2}) we can see that the correlation function
directly measures the spacetime correlations within the source.

It may at first seem that all of the components of the correlation
tensor $B^{-1}$ can be found simply by comparing the results of a
four-dimensional fit to the correlation function with Eq.~(\ref{1.4}).
Such a fit is not possible, however, since $q_0$ is highly correlated
with the other components of ${\bf q}$ through the equation
\begin{equation}
   q_0 = \sum_i\beta_i q_i\;,\;\;\;\;\;\;{\rm where}\;\;\;\;\;\;\;
   \beta_i = \frac{2 K_i}{E_1+E_2}\;.
 \label{1.7}
 \end{equation}
By making the approximation
 \begin{equation}
   \beta_i \simeq  K_i/E_K\;,
 \label{1.8}
 \end{equation}
which is valid for pairs with $|{\bf q}|\ll E_K$, one can use
(\ref{1.7}) to fit the correlation function (\ref{1.4}) to the
form
 \begin{eqnarray}
   C({\bf q},{\bf K}) =  1\pm \lambda\exp
   &\Bigl[&- q_1^2 R_1^2({\bf K})
     - q_2^2 R_2^2({\bf K}) - q_3^2 R_3^2({\bf K})
 \nonumber \\
   &&- 2 q_1 q_2 R_{12}^2({\bf K}) - 2 q_1 q_3 R_{13}^2({\bf K})
     - 2 q_2 q_3 R_{23}^2({\bf K}) \Bigr]\;,
 \label{1.9}
 \end{eqnarray}
where the $R_{ij}^2$ cross terms can be either positive or negative
and $\lambda$ is a parameter introduced to allow for
coherence effects \cite{gyul,wein} and/or particles from the decay of
long-lived resonances \cite{bolz,csor1,fields,sull}. From (\ref{rem2}) and
(\ref{1.7}) the six functions $R_{ij}^2({\bf K})$ ($R_i \equiv R_{ii}$)
can be expressed as the correlations \cite{chap2,chap3}
 \begin{equation}
   R_{ij}^2({\bf K}) = \langle (x_i - \beta_i t)(x_j - \beta_j t)\rangle
    - \langle x_i - \beta_i t \rangle \langle x_j - \beta_j t\rangle
  \, .
 \label{1.9a}
\end{equation}
In general, the six $R^2$ parameters found by fitting correlations to
Eq.~(\ref{1.9}) do not provide enough information to determine the 10
independent components of $(B^{-1})_{\mu\nu}$.

Due to Eqs.~(\ref{1.7}) and (\ref{1.8}), any Lorentz transformation
and/or spatial rotation to a new coordinate system can be written as a
purely spatial linear transformation,
\begin{equation}
   q_i^{\prime} = a_{ij} q_j\;.
 \label{1.10}
 \end{equation}
Thus the new $R^{\prime 2}$ parameters found in the primed frame will
simply be linear combinations of the $R^2$ parameters found in the
original frame. For example, the longitudinally co-moving system
(LCMS) is defined as the longitudinally boosted frame in which
$\beta_3^{\prime}=0$ \cite{both,lcms,nachtrag}. The $R^{\prime 2}$
parameters in this (primed) frame are related to those in some fixed
(unprimed) frame via
 \begin{eqnarray}
   R_1^{\prime 2} &=& R_1^2 +\gamma^4 \beta_1^2\beta_3^2 R_3^2
   +2\gamma^2\beta_1\beta_3 R_{13}^2
 \nonumber \\
   R_2^{\prime 2} &=& R_2^2 +\gamma^4 \beta_2^2\beta_3^2 R_3^2
   +2\gamma^2\beta_2\beta_3 R_{23}^2
 \nonumber \\
   R_3^{\prime 2} &=& \gamma^2 R_3^2
 \nonumber \\
   R_{12}^{\prime 2} &=& R_{12}^2 + \gamma^2\beta_2\beta_3 R_{13}^2
   +\gamma^2\beta_1\beta_3 R_{23}^2 + \gamma^4\beta_1\beta_2\beta_3^2 R_3^2
 \nonumber \\
   R_{13}^{\prime 2} &=& \gamma R_{13}^2 +\gamma^3\beta_1\beta_3 R_3^2
 \nonumber \\
   R_{23}^{\prime 2} &=& \gamma R_{23}^2 +\gamma^3\beta_2\beta_3 R_3^2
 \;,\label{1.11}
\end{eqnarray}
where the $\beta_i$ are measured in the fixed frame and $\gamma =
1/\sqrt{1-\beta_3^2}$.  The above equalities can be used as an
experimental test of the validity of the saddle point approximation in
the following way: If measured LCMS radii are not equal to the above
corresponding combinations of fixed-frame radii, then the saddle point
formalism is not a good approximation to the actual correlation
function.

For sources which have highly non-gaussian spacetime dependencies, it
is better to define \cite{WSH95} the inverse of the curvature tensor
directly through Eqns.~(\ref{rem2}) and (\ref{1.6a}) (using the full
source $S(x,K)$) rather than to use second derivatives to define the
curvature tensor at the saddle point via (\ref{1.3}). A simple
example of when this is necessary is provided by the emission function
from a uniform sphere:
\begin{equation}
   S(x,K) = f(K)\delta(t-t_0)\theta(R-r)\;,
 \label{1.12}
\end{equation}
where $r=\sqrt{x^2+y^2+z^2}$ and $R$ is the radius of the sphere. For
this source function, the saddle point is not unique and second
derivatives are ill-defined. Nevertheless, $(B^{-1})_{\mu\nu}$ in
(\ref{rem2}) is perfectly well defined, producing through (\ref{1.4})
a gaussian correlation function of the form
\begin{equation}
   C({\bf q},{\bf K}) = 1 \pm \exp\left(-|{\bf q}|^2R^2/5\right)\;.
 \label{1.13}
 \end{equation}
The reader can verify that the above expression is a good
approximation to the correlation function
 \begin{equation}
   C({\bf q},{\bf K}) = 1 \pm 9
   \left[\frac{\cos(|{\bf q}|R)}{(|{\bf q}|R)^2}
   -\frac{\sin(|{\bf q}|R)}{(|{\bf q}|R)^3}\right]^2\;,
 \label{1.14}
 \end{equation}
which is found by plugging (\ref{1.12}) directly into (\ref{1.0}).
This example clearly shows that in terms of the correlation function,
even a highly non-gaussian source can be well approximated by a
gaussian with the same rms width.

%%%%%%%%%%%%%%%%%%%%%%%%%%%%%%%%%%%%%%%%%%%%%%%%%%%%%%%%%%%%%%%%%%%%%%
\section{Azimuthally Symmetric Sources}\label{s3}
%%%%%%%%%%%%%%%%%%%%%%%%%%%%%%%%%%%%%%%%%%%%%%%%%%%%%%%%%%%%%%%%%%%%%%

For an azimuthally symmetric source, it is convenient to choose
$\hat{z}$ to point along the beam (``longitudinal'') axis and to
choose $\hat{x}$ to point in the same direction as the component of
${\bf K}$ which is perpendicular to the beam (``out'').  The remaining
(``side'') direction is then defined by $\hat{y}=\hat{z}\times
\hat{x}$ \cite{bertsch}. Note that by definition $K_1$ and
$\beta_1$ are always positive and $K_2=\beta_2=0$. Since the latter
is true, azimuthally symmetric sources must satisfy
 \begin{equation}
   S(t,x,y,z,K) = S(t,x,-y,z,K)\;.
 \label{2.1}
\end{equation}
{}From Eq.~(\ref{1.2}) this implies that $\bar{y}=B_{\mu 2}=B_{2\mu}=0$
for $\mu\ne 2$, so for azimuthally symmetric sources $B_{\mu\nu}$ has
only {\it seven} independent components.

By inserting Eq.~(\ref{2.1}) into Eq.~(\ref{1.0}), one can see that
the correlation function from an azimuthally symmetric source must be
unchanged under the substitution $q_2\rightarrow -q_2$.  This implies
that $R_{12}^2=R_{23}^2=0$, so that only four $R^2$ parameters can be
found by making a gaussian fit to the correlation function
\cite{chap2,chap3,chap4}:
 \begin{equation}
   C({\bf q},{\bf K}) =  1\pm \lambda\exp \Bigl[- q_1^2 R_1^2({\bf K})
   - q_2^2 R_2^2({\bf K}) - q_3^2 R_3^2({\bf K})
   - 2 q_1 q_3 R_{13}^2({\bf K}) \Bigr]\;.
 \label{2.2}
 \end{equation}
Expressing these four correlation radii in terms of the seven
independent elements of the curvature tensor we have
\begin{eqnarray}
   R_2^2 &=&\lambda_2^2
\nonumber \\
   R_1^2 &=& \Gamma  \left[ \lambda_1^2 (1-
   \lambda_0^2\lambda_3^2B_{30}^2) +\beta_1^2\lambda_0^2 (1-
   \lambda_1^2\lambda_3^2B_{31}^2)
   +2\beta_1\lambda_1^2\lambda_0^2(B_{10}-\lambda_3^2B_{31}B_{30})
   \right]
 \nonumber \\
   R_3^2 &=& \Gamma \left[ \lambda_3^2 (1-
   \lambda_0^2\lambda_1^2B_{10}^2) +\beta_3^2\lambda_0^2 (1-
   \lambda_1^2\lambda_3^2B_{31}^2)
   +2\beta_3\lambda_3^2\lambda_0^2(B_{30}-\lambda_1^2B_{31}B_{10})
   \right]
 \nonumber \\
   R_{13}^2 &=& \Gamma \Bigl[ \beta_1\beta_3\lambda_0^2 (1-
   \lambda_1^2\lambda_3^2B_{31}^2)
   +\beta_1\lambda_3^2\lambda_0^2(B_{30}-\lambda_1^2B_{31}B_{10})
 \nonumber \\
   & &\;\;\;\;\; +\beta_3\lambda_1^2\lambda_0^2
   (B_{10}-\lambda_3^2B_{31}B_{30})
   -\lambda_1^2\lambda_3^2(B_{31}-\lambda_0^2B_{10}B_{30})\Bigr]\;,
 \label{2.4}
 \end{eqnarray}
where
 \begin{equation}
   \Gamma = \left[1-\lambda_0^2\lambda_1^2B_{10}^2
   -\lambda_1^2\lambda_3^2B_{31}^2-\lambda_0^2\lambda_3^2B_{30}^2
   +2\lambda_0^2\lambda_1^2\lambda_3^2B_{10}B_{30}B_{31}\right]^{-1}
 \label{2.5}\;.
\end{equation}
It is now apparent that the statement that $R_1^2-R_2^2$ is
proportional to the square of the emission time is highly model
dependent.  Implicit in this statement are the assumptions that
$\lambda_1=\lambda_2$ and that all $x$-$z$, $x$-$t$, and $z$-$t$
correlations are negligible.

Until now we have been discussing fits to the correlation function
which are three-dimensional gaussians in the spatial components of
$q$.  It is also possible to fit correlation functions in a different
three-dimensional space defined by $q_\perp=\sqrt{q_1^2+q_2^2}$, $q_3$
and $q_0$.  As we pointed out in the last section, $q_0$ is highly
correlated with the spatial components of $q$ through Eq.~(\ref{1.7})
which we rewrite here as
\begin{equation}
q_0 = \beta_1 q_\perp \cos\phi + \beta_3 q_3\;,
\label{2.5a1}
\end{equation}
where $\phi$ is the angle between $\hat{x}$ and ${\bf q}_\perp$.  Due
to this correlation, $q_0$ only varies over a finite range which is
nonetheless greater than zero for any nonzero values of $\beta_1$ and
$q_\perp$:
\begin{equation}
-\beta_1 q_\perp + \beta_3 q_3 \le q_0 \le \beta_1 q_\perp
+\beta_3 q_3\;.
\label{2.5a2}
\end{equation}
Despite this phase-space limitation, it still possible to fit
correlation functions with gaussians in a $(q_\perp,q_3,q_0)$ space
\cite{vince}.

For example, one could use an azimuthally symmetric generalization of
the Yano-Koonin formalism
\cite{yano,zajc}
 \begin{equation}
   C({\bf q},{\bf K}) =  1\pm \lambda\exp \Bigl[- q_\perp^2 R_t^2
   +(q_0^2-q_3^2)R_4^2 - (q{\cdot}U)^2 (R_0^2+R_4^2) \Bigr]\;,
 \label{2.5a}
 \end{equation}
where
 \begin{equation}
   U = \gamma\,\Bigl(\;1,\;0,\;0,\;v\;\Bigr)\;,
  \;\;\;\;\;\;\;\; \gamma=1/\sqrt{1-v^2}
 \label{2.5b}
\end{equation}
(in units with $c=1$), and $R_t({\bf K})$, $R_0({\bf K})$, $R_4({\bf
K})$, and $v({\bf K})$ are the four fit parameters.  When using this
fitting procedure, it is convenient to define the particle labeling
such that $q_0$ is always positive.  In this way, pairs with positive
and negative $q_3$ are physically distinct and can be separately
binned.  For the remainder of this paper, we will refer to
Eqs.~(\ref{2.2}) and (\ref{2.5a}) as the ``standard'' and ``GYK''
(generalized Yano-Koonin) fitting procedures, respectively.

The boost-invariant form of the ``GYK'' fit means that for any given
value of ${\bf K}$ there is a longitudinally boosted reference frame
in which $v({\bf K})=0$.  The fit parameters $R_t({\bf K})$, $R_0({\bf
K})$, and $R_4({\bf K})$ measure the source in this frame, regardless
of which longitudinally boosted frame is chosen for the evaluation of
the $q_\mu$.  Although the general interpretation of the $v({\bf
K})=0$ frame is difficult, it can be shown that in this frame
 \begin{equation}
   \Bigl\langle (\beta_1 z -\beta_3 x)(\beta_1 t - x)\Bigr\rangle
  -\Bigl\langle \beta_1 z -\beta_3 x \Bigr\rangle
   \Bigl\langle \beta_1 t - x \Bigr\rangle
  -\beta_3 \Bigl\langle y^2 \Bigr\rangle = 0\;.
 \label{2.5c}
 \end{equation}
Furthermore, in this frame the remaining fit parameters take the form
(for $\beta_1\ne 0$)
 \begin{eqnarray}
   R_t^2 &=& \Bigl\langle y^2\Bigr\rangle = \lambda_2^2
 \nonumber \\
   R_4^2 &=& \left\langle \left(z-\frac{\beta_3}{\beta_1}x\right)^2
    \right\rangle
   -\left\langle z-\frac{\beta_3}{\beta_1}x\right\rangle^2
   -\frac{\beta_3^2}{\beta_1^2}\Bigl\langle y^2 \Bigr\rangle
 \nonumber \\
   R_0^2 &=& \left\langle \left(t-\frac{1}{\beta_1}x\right)^2
    \right\rangle
   -\left\langle t-\frac{1}{\beta_1}x\right\rangle^2
   -\frac{1}{\beta_1^2}\Bigl\langle y^2 \Bigr\rangle\;.
 \label{2.6}
\end{eqnarray}
Just as for the correlation radii of (\ref{1.9a}) and (\ref{2.4}),
these fit parameters contain mixtures of various spatial and temporal
lengths of homogeneity, making the extraction of source parameters in
general highly model dependent. However, a ``GYK'' fit has the
advantage that the time structure of the source enters in only one of
the radius parameters, namely $R_0^2$.

%%%%%%%%%%%%%%%%%%%%%%%%%%%%%%%%%%%%%%%%%%%%%%%%%%%%%%%%%%%%%%%
\section{The Measurement Frame}\label{s4}
%%%%%%%%%%%%%%%%%%%%%%%%%%%%%%%%%%%%%%%%%%%%%%%%%%%%%%%%%%%%%%%

In order to illustrate the utility of the saddle-point formalism and
the importance of picking the right longitudinal reference frame for
making measurements, we will now study three example source models.

%%%%%%%%%%%%%%%%%%%%%%%%%%%%%%%%%%%%%%%%%%%%%%%%%%%%%%%%%%%%%%%%%
\subsection{Static Gaussian Source}\label{s4a}
%%%%%%%%%%%%%%%%%%%%%%%%%%%%%%%%%%%%%%%%%%%%%%%%%%%%%%%%%%%%%%%%%

Consider a source function which is defined by spacetime gaussians in
its center-of-mass (CM) frame:
 \begin{equation}
    S(x,K) = f(K)\, \exp\left[-\frac{x^2+y^2}{2R^2} -\frac{z^2}{2L^2}
    -\frac{(t-t_0)^2}{2(\Delta t)^2}\right]\;.
 \label{1.3a}
 \end{equation}
Using Eq.~(\ref{1.1}), we see that the saddle point for this function
is independent of ${\bf K}$ and given by $\bar{x}=\bar{y}=\bar{z}=0$
and $\bar{t}=t_0$.  In the CM frame, all of the off-diagonal
components of $B_{\mu\nu}$ vanish, while the curvature radii are given
by
 \begin{equation}
   \lambda_1=\lambda_2=R\;,\;\;\;\;\;\;
   \lambda_3=L\;,\;\;\;\;\;\;{\rm and}\;\;\;\;\;\;
   \lambda_0=\Delta t\;.
 \label{1.3b}
 \end{equation}
That the curvature radii are independent of ${\bf K}$ is true
for any source in which the spacetime and momentum dependences
factorize. For the simple source of Eq.~(\ref{1.3a}) they simply
measure the relevant geometrical ``radii" of the system.

If one makes a ``standard'' fit to the correlation function using CM
momentum differences, the extracted radii will have the interpretation
 \begin{eqnarray}
   R_1^2 &=& R^2 + \beta_1^2(\Delta t)^2\;,\;\;\;\;\;\;\;\;
   R_2^2 = R^2
 \nonumber \\
   R_3^2 &=& L^2 + \beta_3^2(\Delta t)^2\;,\;\;\;\;\;\;\;\;
   R_{13}^2 = \beta_1\beta_3(\Delta t)^2
 \label{ex.1}
 \end{eqnarray}
The duration of emission $(\Delta t)$ can thus be extracted either
from $R_{13}^2$ or from the difference $R_1^2-R_2^2$.  If some frame
other than the CM is used for making a ``standard'' fit, then the
correlation radii will have much more complicated dependencies on the
source parameters.  In particular, $R_1^2-R_2^2$ will not be
proportional to $(\Delta t)^2$.

If a ``GYK'' fit is made, on the other hand, then in the CM frame or
{\em any} frame longitudinally boosted from it, one will find the same
expressions for the correlation radii:
 \begin{equation}
   R_t = R\;,\;\;\;\;\;\;\; R_4 = L\;,\;\;\;\;\;\;\;
   R_0 = \Delta t\;.
 \label{ex.2}
 \end{equation}
Furthermore, the parameter $v$ will measure the velocity difference
between the CM and measurement frames.  In other words, even if
measurements are made in the ``wrong'' frame, a ``GYK'' fit will still
produce correlation radii which measure the source in its rest frame.

%%%%%%%%%%%%%%%%%%%%%%%%%%%%%%%%%%%%%%%%%%%%%%%%%%%%%%%%%%%%%%%%%
\subsection{Boost-Invariant Source}\label{s4b}
%%%%%%%%%%%%%%%%%%%%%%%%%%%%%%%%%%%%%%%%%%%%%%%%%%%%%%%%%%%%%%%%%

Next we consider a longitudinally expanding, boost-invariant source:
 \begin{equation}
   S(x,K) = \frac{m_t}{(\Delta\tau)}\exp\left[-\frac{K{\cdot}u}{T}
   -\frac{x^2+y^2}{2R^2} -\frac{(\tau-\tau_0)^2}{2(\Delta\tau)^2}
   \right]\;.
 \label{ex.3}
 \end{equation}
Here $T$ is a constant freeze-out temperature, $\tau=\sqrt{t^2-z^2}$
is the longitudinal proper time, and $m_t = \sqrt{m^2 + K_\perp^2}$.
The longitudinal expansion of the source is described by the flow
four-velocity relative to some fixed frame
 \begin{equation}
   u^\mu = \left(\; {\rm ch}\,\eta,\;0,\;0,\;{\rm sh}\,\eta\;\right)\;,
 \label{ex.4}
 \end{equation}
where $\eta = {\textstyle \frac{1}{2}} \ln[(t+z)/(t-z)]$ is spacetime
rapidity.

Calculation of the saddle point in the fixed frame is straightforward,
yielding $\bar{x}=\bar{y}=0$, $\bar{\tau}=\tau_0$ and
$\bar{\eta}=Y$, where $Y$ is the rapidity of a particle
with momentum ${\bf K}$.  Unlike in the previous model, both the
saddle point and the curvature tensor depend on ${\bf K}$.
Explicitly, the curvature radii and off-diagonal elements of
$B_{\mu\nu}$ are given by
 \begin{eqnarray}
   \lambda_0 &=& \left[ \frac{{\rm sh}^2Y}{\tau_0^2}\left(
   \frac{m_t}{T}\right) +\frac{{\rm ch}^2Y}{(\Delta\tau)^2}\right]^{-1/2}
 \nonumber \\
   \lambda_1 &=& \lambda_2 = R
 \nonumber \\
   \lambda_3 &=& \left[ \frac{{\rm ch}^2Y}{\tau_0^2}\left(
   \frac{m_t}{T}\right) +\frac{{\rm sh}^2Y}{(\Delta\tau)^2}\right]^{-1/2}
 \nonumber \\
   B_{03} &=& B_{30} = -{\rm sh}Y\,{\rm ch}Y\left[ \frac{1}{\tau_0^2}\left(
   \frac{m_t}{T}\right) +\frac{1}{(\Delta\tau)^2}\right]
 \label{ex.5}
 \end{eqnarray}
Here we see that the $z$-$t$ correlations of the
longitudinally expanding source give rise to a nonvanishing $B_{30}$
for pairs with $Y\ne 0$.

The correlation radii from a ``standard'' fit can be found by plugging
these source parameters into Eq.~(\ref{2.4}).  One finds
 \begin{eqnarray}
   R_2^2 &=& R^2
 \nonumber \\
   R_1^2 &=& R^2 + \beta_1^2\left[(\Delta\tau)^2{\rm ch}^2Y
   +\left(\frac{T}{m_t}\right)\tau_0^2{\rm sh}^2Y\right]
 \nonumber \\
   R_3^2 &=& \left(\frac{T}{m_t}\right)\frac{\tau_0^2}{{\rm ch}^2Y}
 \nonumber \\
   R_{13}^2 &=& -\beta_1\left(\frac{T}{m_t}\right)\tau_0^2\,{\rm th}Y\;.
 \label{ex.6}
\end{eqnarray}
Using a saddle point approximation to a hydrodynamic model similar to
the one presented here, Makhlin and Sinyukov first derived the above
expression for $R_3^2$ \cite{makl}. In a recent paper by the NA35
collaboration at CERN, the $m_t$ and $Y$ dependence of $R_3$ measured
in relativistic heavy-ion collisions was compared to that expression
in order to estimate a freezeout proper time $\tau_0$ \cite{na35alb}.
This treatment was not consistent, however, since the parameter
$R_{13}^2$ was omitted from the fits.

Notice that $R_1^2-R_2^2$ in Eqs.~(\ref{ex.6}) is in general
proportional neither to $(\Delta\tau)^2$ nor to $\lambda_0^2$.  This
is simply a result of choosing the wrong measurement frame.  The
extraction of $\Delta\tau$ from correlation radii is greatly
simplified if instead of measuring the correlation in a fixed frame,
one measures it in the local rest frame of the source near its saddle
point $\bar{x}({\bf K})$. At that point, the flow velocity relative to
the fixed frame is given by
\begin{equation}
   u = \biggl(\; {\rm ch}\bar{\eta},\;0,\;0,\;{\rm sh}\bar{\eta}\;\biggr)
 \label{ex.7}
 \end{equation}
Since $\bar{\eta}=Y$, transforming to the local rest frame (primed) of
the saddle point can be done for each pair by making the following
substitutions
\[
  \eta^\prime = \eta - \bar{\eta} = \eta - Y
  \;,\;\;\;\;\;\; Y^\prime = Y - \bar{\eta} = 0
\]
 \begin{equation}
   q_3^\prime = {\rm ch}Y(q_3 - {\rm th}Y\,q_0)\;,\;\;\;\;\;\;\;
   q_0^\prime = {\rm ch}Y(q_0 - {\rm th}Y\,q_3)\;.
 \label{ex.8}
 \end{equation}
In other words, the local rest frame of a boost-invariant source is
just the LCMS ($Y^\prime =0$ frame).

It is easy to verify that in this frame $B^\prime_{30}$ vanishes and
 \begin{equation}
   \lambda_0^\prime = \Delta\tau\;,\;\;\;\;\;\;
   \lambda_1^\prime = \lambda_2^\prime = R\;,\;\;\;\;\;\;
   \lambda_3^\prime = \tau_0\sqrt{T/m_t}\;.
 \label{ex.9}
 \end{equation}
Due to the diagonal nature of the curvature tensor, these source
parameters can be easily extracted from the LCMS correlation radii
 \begin{equation}
   R_1^{\prime 2} = R^2 + \beta_1^2(\Delta\tau)^2\;,\;\;\;\;\;\;
   R_2^{\prime 2} = R^2\;,\;\;\;\;\;\;
   R_3^{\prime 2} = (T/m_t)\tau_0^2\;,\;\;\;\;\;\;
   R_{13}^{\prime 2} = 0\;.
 \label{ex.10}
 \end{equation}
Note that in this frame $R_1^2-R_2^2$ {\em is} proportional to
$(\Delta\tau)^2$.  Using the ``standard'' fitting procedure, it
obviously makes the most sense to measure correlations for
boost-invariant sources in the LCMS rather than a fixed frame.

Using the ``GYK'' fitting procedure, on the other hand, one finds that
regardless of the longitudinal frame chosen,
 \begin{equation}
   R_t = R\;,\;\;\;\;\;\;
   R_4 = \tau_0\sqrt{T/m_t}\;,\;\;\;\;\;\;
   R_0 = \Delta\tau\;,
 \label{ex.11}
 \end{equation}
while $v$ measures the difference between the measurement frame and the
local rest frame of the source at the saddle point.  For example,
a measurement made in the LCMS would yield $v=0$, while one made in a
fixed frame would yield $v={\rm th}\,Y$ for pairs with an average
rapidity $Y$ relative to that frame.

%%%%%%%%%%%%%%%%%%%%%%%%%%%%%%%%%%%%%%%%%%%%%%%%%%%%%%%%%%%%%%%%%
\subsection{Finite Expanding Source}\label{s4c}
%%%%%%%%%%%%%%%%%%%%%%%%%%%%%%%%%%%%%%%%%%%%%%%%%%%%%%%%%%%%%%%%%

The main problem with a boost-invariant source is that it gives rise
to a $dN/dy$ which is completely flat, whereas the $dN/dy$ for
produced particles which are actually observed in relativistic
heavy-ion collisions are much better described by gaussians in
rapidity. Furthermore, in a boost-invariant source, $R_{13}^{\prime
2}$ will vanish in the LCMS, but nonzero values for $R_{13}^{\prime
2}$ have been measured by NA35 \cite{alberp}. These inconsistencies
with boost-invariant sources lead us to consider an expanding model
with a non-boost-invariant cutoff in spacetime rapidity
\cite{chap2,chap3,csor2,padu}.
Using the expansion four-velocity of Eq.~(\ref{ex.4}), we define a
source in its CM frame by
\begin{equation}
   S(x,K) = \frac{m_t\, {\rm ch}(\eta-Y)}
            {(2\pi)^3 \sqrt{2\pi(\Delta \tau)^2}}
      \exp \left[- \frac{K{\cdot}u}{T}
                 - \frac{x^2+y^2}{2R^2}
                 - \frac{(\tau-\tau_0)^2}{2(\Delta \tau)^2}
                 - \frac{\eta^2}{2(\Delta \eta)^2}
           \right]\, .
\label{ex.11a}
\end{equation}
In \cite{chap2} it was shown that the $\Delta\eta$ cutoff term leads
to more realistic gaussian-like rapidity distributions. The prefactor
was introduced so that in the limit as $\Delta\tau\to 0$, $S(x,K)$
becomes the Boltzmann approximation to a hydrodynamical source which
freezes out at a constant temperature $T$ and proper time $\tau_0$
\cite{marb}.

The saddle point for this model still has the coordinates
$\bar{x}=\bar{y} = 0$ and $\bar{\tau} = \tau_0$, but the
spacetime-rapidity coordinate for a given value of ${\bf K}$ (relative
to the CM frame) is now given by the solution of
 \begin{equation}
   {\rm th}(\bar{\eta}-Y) - \frac{m_t}{T}\,{\rm sh}(\bar{\eta}-Y) -
   \frac{\bar{\eta}}{(\Delta\eta)^2} = 0\;.
 \label{ex.12}
 \end{equation}
It is apparent that the saddle point is only located at $\bar{\eta}=Y$
for an infinite longitudinal tube $\Delta\eta\rightarrow\infty$ or for
pairs with $Y=0$.

As in the previous model, transforming to the local rest frame
(primed) at the saddle point can be achieved by
 \[
   \eta^\prime = \eta - \bar{\eta}\;,\;\;\;\;\;\;
   Y^\prime = Y - \bar{\eta}
 \]
 \begin{equation}
   q_3^\prime = {\rm ch}\bar{\eta}(q_3 - {\rm th}\bar{\eta}\,q_0)\;,
   \;\;\;\;\;\;\;
   q_0^\prime = {\rm ch}\bar{\eta}(q_0 - {\rm th}\bar{\eta}\,q_3)\;.
 \label{ex.13}
 \end{equation}
Notice that since $Y^\prime\ne 0$, the local rest frame {\em does
not} coincide with the LCMS.  It is possible to show that in the local
rest frame the curvature tensor becomes diagonal ($B_{30}^\prime=0$)
and
 \[
   \lambda_1^\prime = \lambda_2^\prime = R\;,\;\;\;\;\;\;
   \lambda_0^\prime = \Delta\tau\;,
 \]
 \begin{equation}
   \lambda_3^\prime = \tau_0\left[\frac{m_t}{T}{\rm ch}(\bar{\eta}-Y)
   -\frac{1}{{\rm ch}^2(\bar{\eta}-Y)} +\frac{1}{(\Delta\eta)^2}
   \right]^{-1/2}\;.
 \label{ex.14}
 \end{equation}
In the CM and LCMS frames, however, the curvature tensor is in general
not diagonal and each component becomes much more complicated. These
complications carry over into the correlation radii. For example,
$R_{13}^2$ does not vanish in these frames, and $R_1^2-R_2^2$ is not
proportional to either $(\Delta\tau)^2$ or $\lambda_0^2$
\cite{chap2,chap3}.

By looking at one-particle slopes and rapidity distributions, it may
be possible to estimate $T$ and $\Delta\eta$.  Using these values, one
can numerically solve Eq.~(\ref{ex.12}) for $\bar{\eta}({\bf K})$, use
Eq.~(\ref{ex.13}) to transform to the local rest frame, and then
extract the source parameters by making a ``standard'' fit to the
correlation function.  Alternatively, one could simply make a ``GYK''
fit in the CM frame (or any other longitudinally boosted frame), and
the result would be
 \begin{equation}
   R_t = \lambda_1^\prime = \lambda_2^\prime\;,\;\;\;\;\;\;
   R_4 = \lambda_3^\prime\;,\;\;\;\;\;\;
   R_0 = \lambda_0^\prime\;.
 \label{ex.15}
 \end{equation}
Again the parameter $v({\bf K})$ would measure the velocity difference
between the measurement frame and the local rest frame.

Each of the three models discussed above exhibits a different local
rest frame.  If one makes a ``standard'' fit to the correlation
function, it is important to guess the correct reference frame {\em
before} performing the fit in order to extract useful information
about the source.  This implies that one must have some a priori
knowledge about the source before making the fit.  If one makes a
``GYK'' fit, however, there is no need to pick a frame beforehand.  In
all three of the above cases, the fitting procedure itself
automatically chooses the correct frame in which to measure the
source.  This feature of the ``GYK'' fitting procedure is actually
common to a whole class of models we will discuss below.

%%%%%%%%%%%%%%%%%%%%%%%%%%%%%%%%%%%%%%%%%%%%%%%%%%%%%%%%%%%%%%%%%
\subsection{A Class of Models}\label{s4d}
%%%%%%%%%%%%%%%%%%%%%%%%%%%%%%%%%%%%%%%%%%%%%%%%%%%%%%%%%%%%%%%%%

All of the models we have discussed so far belong to a class of source
functions which satisfy the following conditions:
\begin{equation}
   \lambda_1=\lambda_2\;\;\;\;\;\;{\rm and}\;\;\;\;\;\;
   B_{10}=B_{31}=0\;.
 \label{3.1}
\end{equation}
For models in this class, the curvature tensor (\ref{1.3}) takes the
simple block-diagonal form
\begin{equation}
  B_{\mu\nu} = \left(
    \begin{array}{cccc}
       \lambda_2^{-2} &  0  &  0  &  0  \\
        0  & \lambda_2^{-2} &  0  &  0  \\
        0  &  0  & \lambda_3^{-2} & B_{30} \\
        0  &  0  & B_{30} & \lambda_0^{-2}
    \end{array}
   \right) \, .
\label{3.1a}
\end{equation}
Moreover its inverse, the correlation matrix $B^{-1}$ of
Eqs.~(\ref{1.4}) and (\ref{rem2}), has the same block-diagonal
structure, i.e., the $x$-$t$ and $x$-$z$ correlations $(B^{-1})_{10}$
and $(B^{-1})_{13}$ vanish, and $(B^{-1})_{11} = (B^{-1})_{22}$.

Since models in this class have only four non-vanishing components of
the curvature tensor, these four can be unambiguously determined by
measuring the four parameters coming from either a ``standard'' or
``GYK'' fit.  Explicitly,
\begin{eqnarray}
   R_2\; = & \lambda_2 & =\; R_t
 \nonumber \\
   R_3^2 - \frac{(R_{13}^2)^2}{(R_1^2-R_2^2)}\; = & \lambda_3^2 &
   =\; \frac{R_0^2R_4^2}{\gamma^2(R_0^2+v^2R_4^2)}
 \nonumber \\
   \frac{(R_1^2-R_2^2)R_3^2-(R_{13}^2)^2}
   {\beta_1^2 R_3^2 - 2\beta_1\beta_3 R_{13}^2 +
   \beta_3^2(R_1^2-R_2^2)}\; = & \lambda_0^2 &
   =\; \frac{R_0^2R_4^2}{\gamma^2(R_4^2+v^2R_0^2)}
 \nonumber \\
   \frac{\beta_1 R_{13}^2 - \beta_3(R_1^2-R_2^2)}
        {(R_1^2-R_2^2)R_3^2-(R_{13}^2)^2}\; = & B_{30} &
   =\; -\frac{v\gamma^2(R_0^2+R_4^2)}{R_0^2R_4^2}\;,
 \label{3.2}
\end{eqnarray}
where $\gamma$ is defined in (\ref{2.5b}).  If the $q$ used for
fitting the correlation function are evaluated in a fixed frame, then
the above parameters describe the source as seen in that fixed frame.
If, on the other hand, LCMS values for $q_i$ are used, then
Eqs.~(\ref{3.2}) (with $\beta_3=0$) will determine LCMS source
parameters.

An interesting experimental test is provided by comparing the
left and right sides of Eqs.~(\ref{3.2}).  Namely, if after making
both ``standard'' and ``GYK'' fits to the correlation function it is
found that the left and right sides of Eqs.~(\ref{3.2}) are not equal,
then the source in question cannot belong to the class (\ref{3.1}).

As we mentioned previously, the form of the ``GYK'' fit means that
the extracted parameters $R_t({\bf K})$, $R_4({\bf K})$, and $R_0({\bf
K})$ naturally measure the source in the $v({\bf K})=0$ frame.
Setting $v=0$ in Eqs.~(\ref{3.2}), we see that the right halves of the
first three equations reduce to Eqs.~(\ref{ex.15}).  In other words,
the $R$ parameters directly measure the curvature radii of the source
in the $v({\bf K})=0$ frame.  Furthermore, the last equation in
(\ref{3.2}) shows us that the $v({\bf K})=0$ frame corresponds to the
frame in which $B_{30}({\bf K})$ vanishes ($z$-$t$ correlations
vanish) and the curvature tensor diagonalizes.  For many interesting
models including the three discussed above, this frame also
corresponds to the local rest frame of the source at the saddle
point $\bar{x}({\bf K})$.

As for the ``standard'' fitting procedure, the left halves of
Eqs.~(\ref{3.2}) tell us that only if we are clever enough to pick the
$B_{30}=0$ frame beforehand will $\lambda_0^2$
reduce to the more familiar form
 \begin{equation}
   \lambda_0^2 = \frac{1}{\beta_1^2}(R_1^2-R_2^2)\;.
 \label{3.3}
 \end{equation}
Using RQMD events in the LCMS, this equation was recently shown to be
a good approximation only for pairs with very small $Y$ and $K_\perp$
in the center-of-mass frame of a symmetric projectile-target collision
\cite{fields}. The restriction of small $Y$ found by these authors can
be explained in the following way: Only at $Y=0$ do the LCMS and the
local rest frame definitely coincide for all sources resulting from a
symmetric projectile-target collision.  As $|Y|$ increases, these two
frames may begin to diverge just as they did for the model discussed
is subsection \ref{s4c} above.  For large $Y$, $B_{30}$ becomes
non-negligible in the LCMS and (\ref{3.2}) rather than (\ref{3.3})
must be used to extract $\lambda_0$.  As we will show in the next
section, the additional restriction of small $K_\perp$ can be
explained if a source is undergoing transverse as well as longitudinal
expansion.

The reader should note that there is a difficulty with the
``standard'' fitting procedure which arises due to the fact that
experimental correlation functions are always generated for pairs
which lie in certain bins in $K_\perp$ and $K_3{=}K_L$ rather than for
exact values of these average momenta.  In extracting $\lambda_0$ and
$B_{30}$ of Eqs.~(\ref{3.2}) from the left-hand expressions, there
will be a certain ambiguity as to which values of $\beta_1$ and
$\beta_3$ should be used.  Fortunately, this problem can be
circumvented by employing the new fitting procedures that we introduce
in Appendix A.  Since no $\beta_i$ are present on the right side of
Eqs.~(\ref{3.2}), the difficulty does not arise at all when one uses
the ``GYK'' fitting procedure.

Using the generalized Yano-Koonin formalism has the further advantage
that by making a single fit in a fixed frame, one can determine the
source parameters both in that frame via (\ref{3.2}) and {\em
simultaneously} in the $v({\bf K})=0$ frame via (\ref{ex.15}).
Similarly, by making a single fit in the LCMS frame, one can
simultaneously determine both LCMS and $v({\bf K})=0$ source
parameters. Using the ``standard'' fitting procedure, on the other
hand, requires at least two fits (see Appendix A) just to determine
the source parameters in a single frame.

Finally, we would like to discuss the reason why we chose to study the
class of models defined by Eqs.~(\ref{3.1}).  First of all, as we have
seen explicitly, there are a number of interesting models which
naturally fall into this class.  More importantly, however, the source
function for pairs with $K_\perp=0$ from {\em any}
azimuthally symmetric model will always fall into this class.  The
reason for this is simply because when $K_\perp =0$, there is no way
to distinguish between the ``side'' and ``out'' directions.
Consequently, $\lambda_1=\lambda_2$ and since $B_{2\mu}=0$, it must
also be true that $B_{1\mu}=0$.  If all models fall into this class
for pairs with $K_\perp$ exactly vanishing, then there should be a
wide range of models which are ``close'' to being in this class for
pairs with sufficiently small $K_\perp$.  In the next section, we will
look at an illustrative example model.

%%%%%%%%%%%%%%%%%%%%%%%%%%%%%%%%%%%%%%%%%%%%%%%%%%%%%%%%%%%%%%%%%
\section{A Model with Transverse Expansion}\label{s5}
%%%%%%%%%%%%%%%%%%%%%%%%%%%%%%%%%%%%%%%%%%%%%%%%%%%%%%%%%%%%%%%%%

Just as longitudinally expanding sources feature $z$-$t$ correlations
in their CM frame, transversally expanding sources will in general
feature $x$-$t$ correlations in their CM frame or in any frame which
is only longitudinally boosted relative to it.  Consequently, such
sources will exhibit nonvanishing $B_{10}$ and/or $B_{31}$ and thus not
belong to the class (\ref{3.1}).  Nevertheless, by working with
Eqs.~(\ref{2.3}) as well as the corresponding general expressions for
the ``GYK'' fit, it can be shown that if for some range of ${\bf K}$,
all of the $\lambda_\mu$ are of the same order and
\begin{equation}
   (\lambda_1\lambda_0 B_{10})^2 \ll \beta_1^2 \;,\;\;\;\;\;\;
   (\lambda_3\lambda_1 B_{31})^2 \ll \beta_1^2 \;,\;\;\;\;\;\;
   {\rm and}\;\;\;\;\;\;
   1 - \lambda_2^2/\lambda_1^2 \ll \beta_1^2\;,
\label{3.3b}
\end{equation}
then Eqs.~(\ref{3.2}) are still good approximations.  In other words,
Eqs.~(\ref{3.3b}) define a larger class of models for which it is
possible to unambiguously extract the parameters $\lambda_1({\bf
K})\simeq \lambda_2({\bf K})$, $\lambda_3({\bf K})$, $\lambda_0({\bf
K})$, and $B_{30}({\bf K})$.  Furthermore, from Eq.~(\ref{2.5c}) it
can be verified that if Eqs.~(\ref{3.3b}) hold, then in the $v({\bf
K})=0$ frame, $\lambda_0^2\lambda_3^2B_{30}^2\ll 1$, so it is a good
approximation to treat the curvature tensor as being diagonal, and the
correlation radii in this frame are given by Eqs.~(\ref{ex.15}).

We will now examine a specific source model which exhibits transverse
as well as longitudinal expansion to see what kind of restrictions
Eqs.~(\ref{3.3b}) impose on the average momentum ${\bf K}$.  We consider
a source function of the form (\ref{ex.11a}), but with an expansion
four-velocity now given by
\begin{equation}
   u(x) = \Bigl(\,\sqrt{1+(v_t\rho/R)^2}\,{\rm ch}\eta, \,(v_tx/R),
                \,(v_ty/R),\,\sqrt{1+(v_t\rho/R)^2}\,{\rm sh}\eta\,
          \Bigr)\;.
 \label{3.16}
\end{equation}
Since $u(x)$ is well defined and $u{\cdot}u=1$ for arbitrarily large
$v_t$, Eq.~(\ref{3.16}) can be used for modeling relativistic
($v_t\mathrel{\lower.9ex\hbox{$\stackrel{\displaystyle >}{\sim}$}}1$)
as well as nonrelativistic $(v_t\ll 1)$ transverse expansions.  (Only
for nonrelativistic expansions does the parameter $v_t$ represent
the transverse velocity of the source at $\rho=R$.)

By using Eq.~(\ref{1.1}) to calculate the saddle point of the emission
function and recalling that ${\bf K}$ does not have any component in
the $\hat{y}$ (``side'') direction, it is easily found that
$\bar{\tau}=\tau_0$ and $\bar{y}=0$.  Similarly, one can show that
 \begin{equation}
   \frac{v_t \bar{x}}{R} = \frac{ K_\perp v_t^2/T}{1 + (m_t v_t^2/T)
   {\rm ch}(\bar{\eta}-Y) [1+(v_t \bar{x}/R)^2]^{-1/2}}
 \label{b.3}
 \end{equation}
and
 \begin{equation}
   1 - \frac{\lambda_2^2}{\lambda_1^2} = \frac{
   (v_t\bar{x}/R)^2(m_tv_t^2/T){\rm ch}(\bar{\eta}-Y)}
   {(m_tv_t^2/T){\rm ch}(\bar{\eta}-Y)[1+(v_t \bar{x}/R)^2]
   \,+\,\left[1+(v_t \bar{x}/R)^2\right]^{3/2}}\;.
 \label{b.4}
 \end{equation}
{}From these equations, it can be seen that if we demand that
\begin{equation}
   E_K < \frac{T}{v_t^2}\;,
 \label{3.13}
\end{equation}
then $1-\lambda_2^2/\lambda_1^2$ will always be less than $\beta_1^2$.
Actually, given the form of Eqs.~(\ref{b.3}) and (\ref{b.4}), we
are justified in deducing that Eq.~(\ref{3.13}) implies
$1-\lambda_2^2/\lambda_1^2 \ll \beta_1^2$ rather than the weaker
condition $1-\lambda_2^2/\lambda_1^2 <\beta_1^2$.  In Appendix B we
also show that if (\ref{3.13}) is satisfied, then $(\lambda_1\lambda_0
B_{10})^2 \ll \beta_1^2$ and $(\lambda_3\lambda_1 B_{31})^2
\ll \beta_1^2$.  Consequently, in the context of the present
model, Eq.~(\ref{3.13}) is a sufficient condition to justify the use
of the expressions in the last section.

In particular, if one makes a ``GYK'' fit to the correlation function
of the model under consideration, then for pairs satisfying
(\ref{3.13}), the extracted correlation radii will measure the source
in the local {\em longitudinal} rest frame at the saddle point.
Explicitly,
 \begin{eqnarray}
   R_0 \simeq \lambda_0 \simeq \Delta\tau\;,\;\;\;\;\;\;\;
   R_t \simeq \lambda_1 \simeq \lambda_2 \simeq
   R\left[1+\frac{m_tv_t^2}{T}{\rm ch}(\bar{\eta}-Y)\right]^{-1/2}
 \nonumber \\
   R_4 \simeq \lambda_3 \simeq
   \tau_0\left[\frac{m_t}{T}{\rm ch}(\bar{\eta}-Y)
   -\frac{1}{{\rm ch}^2(\bar{\eta}-Y)} +\frac{1}{(\Delta\eta)^2}
   \right]^{-1/2}\;,
 \label{3.13a}
 \end{eqnarray}
where $\bar{\eta}({\bf K})\simeq{\rm th}^{-1}[v({\bf K})]$.  By
looking at the $m_t$-dependence of $R_t$, it may be possible to
extract both the transverse size $R$ and the expansion parameter
$v_t^2/T$.  Using the latter, it is possible to make a consistency
check to see if the pairs under consideration did in fact satisfy
condition (\ref{3.13}).

The temperature $T$ and transverse velocity parameter $v_t$ can also
be determined by measuring slopes and curvatures of one-particle
distributions.  In heavy-ion collisions at the both the AGS and SPS,
these parameters have been estimated to be on the order of $T=140$ MeV
and $v_t=0.5$ \cite{schn,bm,jacak}.  From Eq.~(\ref{3.13}), these
estimates imply that Eqs.~(\ref{3.13a}) should be good approximations
for pairs with $E_K$ less than about 560 MeV\@.

%%%%%%%%%%%%%%%%%%%%%%%%%%%%%%%%%%%%%%%%%%%%%%%%%%%%%%%%%%%%%%%%
\section{Conclusions}\label{s6}
%%%%%%%%%%%%%%%%%%%%%%%%%%%%%%%%%%%%%%%%%%%%%%%%%%%%%%%%%%%%%%%%

We have shown that in general the number of parameters needed to
describe a source in the quadratic saddle-point approximation exceeds
the number of parameters which can be determined by making gaussian
fits to two-particle correlation data. However, we have identified a
wide class of interesting models for which the source is fully
described by only ${\bf K}$-dependent parameters which can all be
determined from the experimental parameters measured in a gaussian
fit. Using a realistic three-dimensionally expanding hydrodynamical
model, we showed that for heavy-ion collisions at the AGS or SPS, it
should be simplest to extract the source parameters from correlations
of pairs with average energies less than about 560 MeV in the
measurement frame. It should be noted that the source shape seen by
these particles may still not be the geometrical shape of the source,
but rather that of the local region of homogeneity which is affected
by the expansion flow profile of the source. To separate the flow
effect from the underlying geometry, it is necessary to determine the
$K_\perp$ dependence of the HBT radius parameters. Fortunately, much
of this dependence may be possible to see while staying in the
``simple'' regime. For example, for pions with $K_L=0$ in the
measurement frame, values of $K_\perp$ up to 540 MeV/c will still
correspond to energies below 560 MeV, so at least five 100-MeV bins in
$K_\perp$ can be explored below the limit.  Of course for any given
analytic model it should also be possible to extract source parameters
from large $K_\perp$ correlation radii by using expressions such as
those in Eqs.~(\ref{2.4}).  However, at large $K_\perp$, each
correlation radius will contain contributions from a large number of
effects which may be difficult to disentangle.

In the past there has been some debate as to which longitudinal
reference frame would be the most appropriate for measuring
correlations from a given reaction.  For example, for a source which
is not expanding longitudinally, the source center-of-mass frame is
the natural choice.  On the other hand, for an infinite source which
is undergoing a boost-invariant expansion, the LCMS represents the
local rest frame of the source and is thus the natural choice.  Since
experimental reactions undoubtedly produce sources which lie somewhere
between these two extremes, some intermediate frame is needed.  The
generalized Yano-Koonin fitting procedure of Eq.~(\ref{2.5a}) has the
advantage that it does not require one to postulate a reference frame
beforehand; the data themselves determine a frame for each value of
${\bf K}$.  For many interesting ``intermediate'' models, the
parameters in this $v({\bf K})=0$ frame measure the source in its
local longitudinal rest frame, while fixed frame or LCMS parameters
measure the source in some different frame.

Certainly the best way to compare any given model to correlation data
is to make the comparison directly in a six-dimensional $({\bf p}_1,{\bf
p}_2)$ space, rather than to compare the fitted correlation radii of the
model to those extracted from the data.  Nevertheless, we have shown here
that gaussian fits can still reveal some very interesting information
about the velocity of the local longitudinal rest frame as well as
the lengths of homogeneity of the source.

\acknowledgments

The authors would like to thank Pierre Scotto, Urs Wiedemann, Doug
Fields and Nu Xu for fruitful discussions and constructive comments.
The work of U.H. was supported by the Deutsche Forschungsgemeinschaft
(DFG), Gesellschaft f\"ur Schwerionenforschung (GSI) and
Bundesministerium f\"ur Bildung und Forschung (BMBF). The work of S.C.
and J.R.N. was supported by the US Department of Energy.

% Here begins the Appendix:

%%%%%%%%%%%%%%%%%%%%%%%%%%%%%%%%%%%%%%%%%%%%%%%%%%%%%%%%%%%%%%%%
\appendix
\section{}
%%%%%%%%%%%%%%%%%%%%%%%%%%%%%%%%%%%%%%%%%%%%%%%%%%%%%%%%%%%%%%%%

In this appendix we show how the $\beta_i$ dependencies in $\lambda_0$
and $B_{30}$ of Eqs.~(\ref{3.2}) can be removed through the
introduction of new fitting procedures.  We label the fitting
procedure defined by Eq.~(\ref{2.2}) with an ``$a$''.  Thus,
$R_3^2(a)$ refers to the square of the ``longitudinal'' radius as
found by fitting the correlation with Eq.~(\ref{2.2}).  We define a
``$b$'' fitting procedure by using $\beta_1 q_3$ instead of $q_3$ in
making a gaussian fit to the correlation function.  In other words,
 \begin{equation}
   C({\bf q},{\bf K}) =  1\pm \lambda\exp \Bigl[- q_1^2 R_1^2(b)
   -q_2^2 R_2^2(b) - (\beta_1 q_3)^2 R_3^2(b)
   - 2 q_1 (\beta_1 q_3) R_{13}^2(b) \Bigr]\;,
 \label{3.4}
 \end{equation}
where we have suppressed the ${\bf K}$ dependence of the $R^2$
parameters.  It can now be seen that in the LCMS (primed) frame
 \begin{eqnarray}
   \lambda_0^{\prime 2} &=&
   \frac{\left[R_1^{\prime 2}(a)-R_2^{\prime 2}(a)\right]
         R_3^{\prime 2}(a)-\left[R_{13}^{\prime 2}(a)\right]^2}
        {R_3^{\prime 2}(b)}
 \nonumber \\
   B_{30}^{\prime} &=&
   \frac{R_{13}^{\prime 2}(b)}
        {\left[R_1^{\prime 2}(a)-R_2^{\prime 2}(a)\right]
         R_3^{\prime 2}(a)-\left[R_{13}^{\prime 2}(a)\right]^2}\;.
 \label{3.5}
 \end{eqnarray}

To calculate $\lambda_0$ and $B_{30}$ in a fixed frame, we need to
introduce two additional fitting procedures.  Procedure ``$c$'' is
defined by using $\beta_3 q_1$, $\beta_3 q_2$, and $\beta_1 q_3$ in
place of $q_1$, $q_2$, and $q_3$ when making gaussian fits, while
procedure ``$d$'' is defined by using $\sqrt{|\beta_3|}q_1$,
$\sqrt{|\beta_3|}q_2$, and $q_3$.  We then have
 \begin{eqnarray}
   \lambda_0^2 &=&
   \frac{\left[R_1^2(a)-R_2^2(a)\right]R_3^2(a)
         -\left[R_{13}^2(a)\right]^2}
        {R_3^2(c) - 2R_{13}^2(c)
        + R_1^2(c)-R_2^2(c)}
 \nonumber \\
   B_{30} &=&
   \frac{R_{13}^2(b) \mp \left[R_1^2(d)-R_2^2(d)\right]}
        {\left[R_1^2(a)-R_2^2(a)\right]R_3^2(a)
         -\left[R_{13}^2(a)\right]^2}\;,
 \label{3.6}
 \end{eqnarray}
where the $-$ ($+$) sign in $B_{30}$ refers to bins in which
$\beta_3>0$ ($\beta_3<0$).  Note that due to this distinction, the
above method for determining source parameters should not be applied
to bins in which some of the pairs have $\beta_3>0$ while others have
$\beta_3<0$.

%%%%%%%%%%%%%%%%%%%%%%%%%%%%%%%%%%%%%%%%%%%%%%%%%%%%%%%%%%%%%%%%
\section{}
%%%%%%%%%%%%%%%%%%%%%%%%%%%%%%%%%%%%%%%%%%%%%%%%%%%%%%%%%%%%%%%%

To prove the remaining inequalities in Eq.~(\ref{3.3b}), we begin by
presenting the following easily verified relations:
\begin{eqnarray}
   B_{31} &=& \frac{m_tv_t}{RT}\left(\frac{v_t\bar{x}}{R}\right)
   \left(\frac{\bar{t}}{\tau_0^2}\right)
   \frac{{\rm sh}(\bar{\eta}-Y)}{\sqrt{1+(v_t \bar{x}/R)^2}}
\nonumber \\
   B_{10} &=& -\frac{m_tv_t}{RT}\left(\frac{v_t\bar{x}}{R}\right)
   \left(\frac{\bar{z}}{\tau_0^2}\right)
   \frac{{\rm sh}(\bar{\eta}-Y)}{\sqrt{1+(v_t \bar{x}/R)^2}}
\nonumber \\
   \frac{1}{\lambda_1^2} &>&
   \frac{m_tv_t^2}{R^2T}\,\frac{{\rm ch}(\bar{\eta}-Y)}
   {[1+(v_t \bar{x}/R)^2]^{3/2}}\;.
 \label{b.4a}
\end{eqnarray}
The saddle point in spacetime rapidity for this source is given by
 \begin{equation}
   {\rm th}(\bar{\eta}-Y) - \frac{m_t}{T}\sqrt{
  1+(v_t \bar{x}/R)^2}\,{\rm sh}(\bar{\eta}-Y)
   -\frac{\bar{\eta}}{(\delta\eta)^2} =0\;.
 \label{3.19}
 \end{equation}
To derive the needed expressions for $\lambda_0^2$ and $\lambda_3^2$,
we have first used Eq.~(\ref{3.19}) to check numerically that if
$(m_t/T)\sqrt{1+(v_t \bar{x}/R)^2}\,>0.7$, then for $Y\ne 0$,
$|\bar{\eta}|<|Y|$ and $\bar{\eta}$ has the same sign as $Y$.
Although $(m_t/T)\sqrt{1+(v_t \bar{x}/R)^2}\,>0.7$ may present a
significant restriction for electron or photon correlation
measurements, it does not present a significant restriction for
current two-hadron correlation measurements from particle or heavy-ion
collisions.  Even for pions, temperatures of up to 200 MeV would still
satisfy this condition.  Given this condition, from Eq.~(\ref{3.19})
it is possible to show that for $Y\ne 0$,
\begin{equation}
   \frac{1}{{\rm ch}(\bar{\eta}-Y)} < \frac{m_t}{T}
   \sqrt{1+(v_t \bar{x}/R)^2}\;.
 \label{b.5}
 \end{equation}
Using this inequality, then for all $Y$ it is true that
 \begin{eqnarray}
   \lambda_0^{-2} &>& \frac{m_t}{T}\sqrt{1+(v_t \bar{x}/R)^2}\,
  \left(\frac{\bar{z}}{\tau_0^2}\right)^2\,
   \frac{{\rm sh}^2(\bar{\eta}-Y)}{{\rm ch}(\bar{\eta}-Y)}
 \nonumber \\
   \lambda_3^{-2} &>& \frac{m_t}{T}\sqrt{1+(v_t \bar{x}/R)^2}\,
  \left(\frac{\bar{t}}{\tau_0^2}\right)^2\,
   \frac{{\rm sh}^2(\bar{\eta}-Y)}{{\rm ch}(\bar{\eta}-Y)}\;.
 \label{b.6}
 \end{eqnarray}
The desired inequalities are now easily proven:
 \begin{eqnarray}
   (\lambda_1\lambda_3B_{31})^2 &<& (v_t\bar{x}/R)^2 \ll \beta_1^2
 \nonumber \\
   (\lambda_1\lambda_0B_{10})^2 &<& (v_t\bar{x}/R)^2 \ll \beta_1^2
 \;.
 \label{b.7}
 \end{eqnarray}

%%%%%%%%%%%%%%%%%%%%%%%%%%%%%%%%%%%%%%%%%%%%%%%%%%%%%%%%%%%%%%%%

%%%%%%%%%%%%%%%%%%%%%%%%%%%%%%%%%%%%%%%%%%%%%%%%%%%%%%%%%%%%%%%%


\begin{thebibliography}{99}

\bibitem{hbt}
  D. Boal, C.K. Gelbke, and B. Jennings, Rev. Mod. Phys. {\bf 62}, 553
  (1990).

\bibitem{na35alb}
  NA35 Coll., T. Alber et al., Z. Phys. C{\bf 66}, 77 (1995).

\bibitem{fixed}
  NA35 Coll., G. Roland et al., Nucl. Phys. A{\bf 566}, 527c (1994);
  NA35 Coll., T. Alber et al., Phys. Rev. Lett. {\bf 74}, 1303 (1995);
  NA44 Coll., M. Sarabura et al., Nucl. Phys. A{\bf 544}, 125c
  (1992);
  E802 Coll., T. Abbott et al., Phys. Rev. Lett. {\bf 69},
  1030 (1992).

\bibitem{both}
  NA35 Coll., P. Seyboth et al., Nucl. Phys. A{\bf 544}, 293c (1992);
  NA35 Coll., D. Ferenc et al., Nucl. Phys. A{\bf 544}, 531c (1992).

\bibitem{lcms}
  NA44 Coll., H. Beker et al., Phys. Rev. Lett. {\bf 74}, 3340 (1995).

\bibitem{sinyu1}
  Yu.M. Sinyukov, in ``Hot Hadronic Matter: Theory and Experiment,"
  edited by J. Letessier et al. (Plenum, New York, 1995).

\bibitem{akke}
  S.V. Akkelin and Yu.M. Sinyukov, Bogolyubov preprint ITP-63-94E, Phys.
  Lett. B, in press.

\bibitem{chap2}
  S. Chapman, P. Scotto, and U. Heinz, Los Alamos e-print archive
  hep-ph/9409349, Heavy Ion Physics {\bf 1}, 1 (1995).

\bibitem{yano}
  F. Yano and S. Koonin, Phys. Lett. B{\bf 78}, 556 (1978).

\bibitem{zajc}
  W. Zajc, Nucl. Phys. A{\bf 525}, 315c (1991).

\bibitem{pratt}
  S. Pratt, T. Cs\"org\H o, and J. Zim\'anyi, Phys. Rev. C{\bf 42}, 2646
  (1990).

\bibitem{chap1}
  S. Chapman and U. Heinz, Phys. Lett. B{\bf 340}, 250 (1994).

\bibitem{bert1}
  G. Bertsch, P. Danielewicz, and M. Herrmann, Phys. Rev. C{\bf 49},
  442 (1994); M. Herr\-mann and G. Bertsch, Phys. Rev. C{\bf 51}, 328
  (1995).

\bibitem{chap3}
  S. Chapman, P. Scotto, and U. Heinz, Phys. Rev. Lett. {\bf 74}, 4400
  (1995).

\bibitem{gyul}
  M. Gyulassy, S.K. Kauffmann, and L.W. Wilson, Phys. Rev. C{\bf 20}, 2267
  (1979).

\bibitem{wein}
  R.M. Weiner pointed out in Phys. Lett. B{\bf 232}, 278 (1989) that
  in the case that the source is partially coherent the correlation
  function should in fact not be fitted to a single gaussian, but to a
  sum of gaussians with radius parameters differing by a factor
  $\sqrt{2}$.

\bibitem{bolz}
  J. Bolz, U. Ornik, M. Plumer, B. Schlei, and R. Weiner, Phys. Lett.
  B{\bf 300}, 404 (1993).

\bibitem{csor1}
  T. Cs\"org\H o, B. L{\o}rstad, and J. Zim\'anyi, Lund preprint
  LUNFD6-NFFL-7088-1994, Phys. Lett. B, submitted.

\bibitem{fields} D. Fields, J. Sullivan, B. Jacak, N. Xu, J.
  Simon-Gillo, and H. van Hecke, Phys. Rev. C, submitted.

\bibitem{sull} J. Sullivan, M. Berenguer, B. Jacak, M. Sarabura,
  J. Simon-Gillo, H. Sorge, H. van Hecke, and S. Pratt, Phys. Rev.
  Lett. {\bf 70}, 3000 (1993).

\bibitem{nachtrag}
  T. Cs\"org\H o and S. Pratt, Lund preprint LU TP 91-10,
  in: {\it Proc. of the Workshop on Relativistic Heavy Ion Physics},
  preprint KFKI-1991-28/A, p. 75.

\bibitem{WSH95}
  U.A. Wiedemann, P. Scotto, and U. Heinz, {\it Transverse momentum
  dependence of HBT correlation radii}, Regensburg preprint TPR-95-12
  (Aug. 1995), unpublished.

\bibitem{bertsch}
  G. Bertsch, M. Gong, and M. Tohyama, Phys. Rev. C{\bf 37}, 1896 (1988).

\bibitem{chap4}
  S. Chapman, P. Scotto, and U. Heinz, Talk given at Quark Matter '95
  in Monterey, January 9--13, 1995, Nucl. Phys. A, to be published.

\bibitem{vince}
  Several such fits are made and the phase-space limitations discussed
  in the 1994 University of Michigan Ph.D. thesis of Vince Cianciolo.

\bibitem{makl}
  A.M. Makhlin and Yu.M. Sinyukov, Z. Phys. C{\bf 39}, 69 (1988).
  Yu.M. Sinyukov, Nucl. Phys. {\bf A498}, 151c (1989).

\bibitem{alberp}
  NA35 Coll., Th. Alber, private communication.

\bibitem{csor2} T. Cs\"org\H o, Phys. Lett. B{\bf 347}, 354 (1995).

\bibitem{padu}
  S. Padula and M. Gyulassy, Nucl. Phys. {\bf A498}, 555c (1988).

\bibitem{marb}
  B.R. Schlei, U. Ornik, M. Pl\"umer, and R.M. Weiner, Phys. Lett.
  B{\bf 293}, 275 (1992).

\bibitem{schn}
  E. Schnedermann and U. Heinz, Phys. Rev. Lett. {\bf 69}, 2908
  (1992);
  E. Schnedermann, J. Sollfrank, and U. Heinz, Phys. Rev. C{\bf 48},
  2462 (1993);
  E. Schnedermann and U. Heinz, Phys. Rev. C{\bf 50}, 1675 (1994).

\bibitem{bm}
  P. Braun-Munzinger, J. Stachel, J. Wessels, and N. Xu, Phys. Lett.
  B{\bf 344}, 43 (1995).

\bibitem{jacak}
  B. Jacak, Talk given at Quark Matter '95 in Monterey,
  January 9--13, 1995, Nucl. Phys. A, to be published.

\end{thebibliography}
\end{document}